\documentclass[aps,prb,showpacs,twocolumn,floats]{revtex4}
\usepackage{amssymb}
\usepackage{amsbsy}
\usepackage{amsmath}
\usepackage{graphicx}
\usepackage{amsmath}
\usepackage{times}
\usepackage{color}
\usepackage{setspace}
\usepackage{bm}
\newcommand\bea{\begin{eqnarray}}
\newcommand\eea{\end{eqnarray}}
\newcommand\beq{\begin{equation}}
\newcommand\eeq{\end{equation}}
\newcommand\bib{\bibitem}

\begin{document}

\title{Two rate periodic protocol with dynamics driven through many cycles}

\author{Satyaki Kar}

\affiliation{Theoretical Physics Department, Indian Association for
the Cultivation of Science, Jadavpur, Kolkata-700032, India.}

\date{\today}

\begin{abstract}

We study the long time dynamics in closed quantum systems  periodically driven 
via time dependent parameters with two frequencies
$\omega_1$ and $\omega_2=r \omega_1$. Tuning of the ratio $r$ there can unleash plenty of dynamical phenomena to occur. 
Our study includes integrable models like
Ising and XY models in $d=1$ and  Kitaev model in $d=1$ and $2$ and can also be extended to Dirac fermions in graphene.
We witness the wave-function overlap or
dynamic freezing to occur within some small/ intermediate frequency regimes in the $(\omega_1 ,r)$ plane (with $r\ne0$) 
when the ground state is evolved through single cycle of driving. 
However, evolved states soon become steady with long driving and the freezing scenario gets rarer.
We extend the formalism of adiabatic-impulse approximation for many cycle driving within our two-rate protocol
and show the near-exact comparisons at small frequencies. An extension of the rotating wave approximation is also
developed to gather an analytical framework of the dynamics at high frequencies.
Finally we compute the entanglement entropy in the stroboscopically evolved states within the gapped phases of the system
and observe how it gets tuned with the ratio $r$ in our protocol.
The minimally entangled states are found to fall within the regime of dynamical freezing.
In general, the results indicate that the entanglement entropy 
in our driven short-ranged integrable systems follow genuine non-area law of scaling and show a convergence (with a $r$ dependent pace) towards volume 
scaling behavior as the driving is continued for long time.

\end{abstract}


\maketitle

\section{Introduction}
\label{intro}

Dynamics of closed quantum many body system, driven out of equilibrium has 
become an engaging research field in recent times\cite{pol1,dziar1,dutta1}. Many interesting physics 
such as universal scaling of excitation density during slow quenches\cite{kz1,pol2,ks1,ks2,pol3, sondhi1} through 
a quantum critical point (QCP), quantum coherence\cite{miyashita,gong} or dynamical phase transitions (DPT) leading to 
cusp-like 
features in Lochsmidt echo pattern\cite{mukh1,pol4,dytr1,dytr2,dutta2} 
has already surfaced in the literature. At the same time, experimental 
realization of such closed quantum system has also became possible using trapped ultra-cold atoms in 
optical lattices\cite{lewenstein} and thus enabling hands-on check for the related theories put forward.
Non-equilibrium dynamics of various integrable and non-integrable systems 
of, for example transverse field Ising (TFIM), XY (TFXYM) or Bose-Hubbard type, can be well emulated these 
days.

A linear quench to drive a system out of equilibrium and the corresponding Kibble Zurek (KZ) mechanism\cite{kz1,dziar1} for 
defect production as it passes through a QCP  has been studied for quite some time.
Study of periodic driving, in comparison, is rather new. 
A periodic protocol allows multiple passage of the system through quantum critical 
points exhibiting novel features like stuckelberg interference, dynamical freezing, dynamical 
phase transitions (DPT) or steady periodic states\cite{stu1,dyn1,dyn2,adas1}. 
In most of these studies, single-frequency periodic protocols are used to drive the system and examine the resulting dynamics. 
Recently Kar $et~al.$\cite{kar} considered a two rate periodic protocol where both the energy bias and detuning in the
underlying two level system (TLS) are treated periodically with time. The low frequency regime that they study there, reveals
signatures of dynamical freezing as driving for one complete cycle is performed. Such low frequency freezing was hitherto unobserved
from the single-rate periodic driving examined so far. For a TFXYM, such freezing occurs mostly around the point where both the periodic functions
have same frequencies, though other class of systems like tilted Bose-Hubbard, Kitaev models or Dirac fermions can  show such phenomena in their respective other discrete 
frequency regimes\cite{kar}. In this present paper, we probe such dynamics stroboscopically as the two-rated driving is continued 
for long time. The alteration in the defect profile is witnessed as we scan through small to large values
in the frequency space.
We find that further regimes of freezing to sprout in the intermediate, not so-large frequencies and discuss their evolution as
the driving is continued stroboscopically for longer times. We analyze the correlation functions, magnetization
and also the behavior of entanglement entropy under such long driving.

Unlike a linear or any other aperiodic drive,
a periodic drive, as mentioned before, unleashes rich set of dynamical phenomena\cite{stu1,dyn1,dyn2,adas1,rev1,nori1,ks3}, 
even more so, if driven by two rate protocols\cite{kar}.
Repeated passage through the critical points give the system, initially in its ground
state, ample chances to move to the excited states and experience the 
quantum interference of the probability waves at the crossing/ avoided crossing 
regions. A drive through a number of cycles allows more exposure to such QCP's
 resulting in alteration of the exotic dynamical responses witnessed within a short time.
Particularly, the high degree of dynamics induced freezing\cite{adas1,pekker1,uma1,rev1}  which implies almost 
complete overlap of the initial and final wave-function, is obtained
with a two-rate protocol just by driving for a single-cycle\cite{kar}. Though the phenomenon, when viewed 
stroboscopically, stays alive for a while,
a long driving tone down the temporal fluctuations of the observables making the freezing scenario rarer. 
This paper reports results of a 1D TFXYM where such freezing takes place
for drive frequency $\omega_1$ as small as unity at  $r\sim 1$ ($r=\omega_2/\omega_1$, $\omega_2$ being the other frequency 
in the two-rate protocol).
A scan through larger frequencies sees spreading of such freezing zone around $r=1$ as well as new freezing
patches to sprout at $\omega_1\sim 10$ for $r\sim 0.5$ and $2.0$, within the ranges shown. With continued spreading for larger
$\omega_1$, freezing gradually takes over the frequency space, for $r$ not very small. A small $\omega_2$ makes the actual
period of the drive very large causing measurements after integral time periods corresponding to $\omega_1$ not truly stroboscopic.
However, a long drive brings in the steady states soon where the transient behaviors, due to temporal fluctuations die out 
resulting in less amount of freezing in the system.

This paper also discusses analytic tools that can well describe the dynamics under the drive with a two rate protocol.
The adiabatic-impulse approximation, used initially in Ref.\onlinecite{kar} for similar protocol, is extended for 
stroboscopic long driving and for larger ranges of frequencies. It turns out to capture the final evolved states at small 
frequencies quite well even after passage through many cycles.
At high frequency regime, a rotating wave approximation is a well known formalism for obtaining a faithful description
of the dynamics. Here we extend the idea to incorporate it to the two rate periodic protocols of ours. The dynamics 
followed by such two rate protocols thus can be well accounted for via analytic frameworks both for small and large frequencies.

Walking down the line, estimating quantum entanglement and its time evolution in the evolved states carries huge importance
from the quantum information perspective.
And the bipartite entanglement entropy is a good measure of such entanglement in the quantum system.
In a many body system of size $L^d$ ($d$ being the dimension), the Von-Neuman entropy $S$ of a sub-system of size $l^d$
can be computed from the reduced density matrix $\rho$ as $S = -{\rm Tr}[\rho{\rm ln}\rho]$.
For the two-rate protocol, $r$ can be treated as a tuning parameter for controlling the system entanglement.
We find that the region of dynamic freezing identifies the minimally entangled states. 
The ground state in a gapped system, corresponding to a short-ranged Hamiltonian, 
obeys an area scaling law for the entropy, $i.e.,~~S\sim l^{d-1}$ (For $d=1$, this is called the 
Hastings' theorem\cite{hastings}).
We see that the entanglement entropy $S_n(l)$, after drive through $n$ cycles, shows such area scaling only when driven for a 
few number of cycles (i.e., $n$ small), but a non-area scaling $S\sim l^{\alpha}$ appears with $\alpha>d-1$ for large $n$ values.
This amounts to a cross-over from a short-ranged to long-ranged behavior as the long drive exposes the system, 
originally defined with a nearest neighbor short
range model, to excitations that are long ranged (described by ground-states of some long-ranged
Hamiltonians\cite{sen1}).
A driving for infinitely long time finally relaxes the system towards a generalized Gibbs ensemble (GGE)\cite{pol1}
with entropy scaling exponent $\alpha\rightarrow d$ ($i.e.,$ approaching volume scaling law).
This convergence of $\alpha$ towards $d$, as $l$ is varied, is found to be non-monotonic for small frequencies.
We see that using $r$ as a tuning parameter, we can expedite or retard the pace of such convergence.

The plan of the paper is as follows. 
In section II, we first describe our two-rate periodic protocol and then analyze the time evolution of the system 
wave-function in a integrable fermionic Hamiltonian (Eq.\ref{fermhamden}) following the Schrodinger equation. 
Here within subsection A, we report and discuss the results on defect densities and magnetization and then formulate the analytic approaches
of adiabatic-impulse approximation and Rotating wave approximation
for our two rate protocol showing their validity in the respective slow and large frequency regimes. 
Then within subsection B, we discuss the entanglement generation during periodic driving and discuss various results on entanglement entropy.
Finally in section III, we summarize our results, discuss their experimental implications,
and conclude.

\section{Formulation and results}

In this work, our aim is to study the effect of two-rate periodic drive 
protocols on a class of integrable closed quantum systems. We choose the drive
protocol to involve periodic variation of two parameters of the
system Hamiltonian with frequencies $\omega_1$ and $\omega_2$. Our protocol represents a class of models that can be 
expressed by a general form of free fermionic Hamiltonian $H=  \sum_{\vec k}
\psi_{\vec k}^{\dagger} H_{\vec k} \psi_{\vec k}$, where $\psi_{\vec
k} = (c_{\vec k}, c_{-\vec k}^{\dagger})^T$ is the two component
fermion field, $c_{\vec k}$ denotes fermionic annihilation operator,
and $H_{\vec k}$ is a $2 \times 2 $ matrix Hamiltonian density:
\begin{eqnarray}
H_{\vec k} &=& \left[ \lambda_1(\omega_1 t) - b_{\vec k}\right]
\tau_3 + \lambda_2(r \omega_1 t) \Delta_{\vec k} \tau_1.
\label{fermhamden}
\end{eqnarray}
Here $\lambda_1(\omega_1t)$ and $\lambda_2(r\omega_1t)$ are two time-dependent parameters with periodic variations given by frequencies 
$\omega_1$ and $\omega_2=r\omega_1$ respectively and $\tau_3 $ and $\tau_1$ are Pauli matrices in particle-hole
space. Such free fermionic Hamiltonians represents Ising and XY
models in $d=1$ \cite{subir0} and Kitaev model in $d=2$
\cite{kit1,sen2,feng1}. 
Additionally, it can also describe singlet/ triplet superconductors and Dirac-like
quasiparticles in graphene \cite{grarev1} and on the surface of a
topological insulators \cite{toprev1}.
In what follows we shall
carry out numerical analysis of this model in the context of XY
model in $d=1$; we note, however, that our results shall be valid
for any other representations of $H_{\vec k}$.

The XY model in a transverse field constitute a model for of
half-integer spins on a one-dimensional (1D) chain having a Hamiltonian
\begin{eqnarray}
H_{\rm XY} = \sum_{\langle i j\rangle, \alpha=x,y} J_{\alpha}
S_i^{\alpha} S_j^{\alpha} - h \sum_i S_i^z.  \label{xyham1}
\end{eqnarray}
Here $J_{x,y}$ are nearest neighbor coupling between $x$ and the $y$
components of the spins, $\langle ij \rangle$ indicates the nearest neighbor
sites $i$ and $j$ and $h$ is the transverse
field. A Jordan-Wigner transformation \cite{dutta1} can map this
Eq.\ \ref{xyham1} to Eq.\ \ref{fermhamden} in $d=1$ with the
identification
\begin{eqnarray}
b_k &=& (J_x + J_y) \cos(k), \quad \lambda_1= -h
\nonumber\\
\Delta_k &=& - i\sin(k), \quad \lambda_2= (J_x-J_y) \label{mapxy}
\end{eqnarray}
So in the two-rate protocol considered, all of the parameters $h,~J_x$ and $J_y$ need to be time-periodic.

In a similar manner, the Kitaev model in $d = 2$ can also be
mapped into Eq.\ref{fermhamden}. The Hamiltonian of the Kitaev model can
be written in terms of half-integer spins residing on the sites
of a honeycomb lattice as 
\begin{align}
 H_{k2D}=\hspace{-.2 in}\sum_{j+l=even}\hspace{-.1 in}(J_1S^x_{j+1,l}S^x_{j,l}+J_2S^y_{j-1,l}S^y_{j,l}+J_3S^z_{j,l+1}S^z_{j,l})
 \label{kit2d}
\end{align}
where $(j,l)$ denotes coordinates of a site on the honeycomb
lattice, and $J_{x,y,z}$ are the coupling strength between neigh-
boring $x,y,z$ components of the spins. It turns out that $H_{k2D}$
can also be mapped to Eq.\ref{fermhamden} with the identification 
\begin{eqnarray}
b_{\vec k}& = &-J_1cos(k_1)-J_2cos(k_2),~~~~~\lambda_1=J_3\nonumber\\
&&\lambda_2\Delta_{\vec k}=J_1sin(k_1)-J_2sin(k_2),
\label{kiteq}
\end{eqnarray}
where $k_{1(2)}= {\vec k}.{\vec M}_{1(2)},~{\vec M}_{1(2)} =\sqrt{3}{\hat i}/2+(-){\hat j}/2$ are the
spanning vectors of the reciprocal lattice and ${\hat i}$ and ${\hat j}$ are the
unit vectors in the $x$ and the $y$ directions.

We can make $J_1=0$ in Eq. \ref{kit2d} to obtain a Kitaev model in one dimension.

Starting from the many-body ground state, the stroboscopic dynamics displays
interesting features in the defect density, magnetization or entanglement profiles.
We give a detailed description of the same in the following.

\subsection{Defect production}

\subsubsection{Numerical integration}

In this section we will describe the time evolution of the system's wave-function following the Schrodinger equation.
In the Jordon-Wigner transformed Hamiltonian, there are $L^d/2$ number ($L^d$ denotes the system size in $d$ dimensions)
of independent ${\vec k}$-modes (involving (${\vec k},-{\vec k}$) pair) due to integrability of the system. 
The two-dimensional subspace spanned by the unoccupied $|0_{+{\vec k}}0_{-{\vec k}}>$ and 
doubly occupied $|{\vec k},-{\vec k}>$ fermionic states, at each ${\vec k}$ level, thus
constitutes a TLS in this problem\cite{dutta1,adas1}.
In the time-independent diabatic basis\cite{nori1}, the instantaneous ground states and excited states for the 
${\vec k}$-mode are
given by $|\phi_{-,\vec k}\rangle = (\beta_{+,\vec k}(t), \beta_{-,\vec k}(t))^T$
and $|\phi_{+,\vec k}\rangle = (\beta_{-,\vec k}(t), -\beta_{+,\vec k}(t))^T$
respectively, with
\begin{eqnarray}
\beta_{\pm,\vec k}(t) &=& \sqrt{\frac{ E_{+,\vec k}\pm(\lambda_1(\omega_1t)-b_{\vec k})}{2E_{+,\vec k}}}, \nonumber\\
E_{\pm,\vec k}(t) &=& \pm\sqrt{(\lambda_1(\omega_1t)-b_{\vec k})^2 +
\lambda_2^2(r\omega_1t)\Delta_{\vec k}^2}. 
\label{gap}
\end{eqnarray}
A general wave-function at a
given momentum $\vec k$ is denoted by
$|\psi_{\vec k}\rangle = (a_{1,\vec
k}(t), a_{2,\vec k}(t))^T$ on the same basis. Now, $b_{\pm,\vec k}$
being the probability amplitudes of the state in the eigen-basis $|\phi_{\pm,\vec k}>$ (or, the time dependent
adiabatic basis), we can write 
\begin{eqnarray}
 a_{1,\vec k}(t)=b_{+,\vec k}(t)\beta_{-,\vec k}(t)+b_{-,\vec k}(t)\beta_{+,\vec k}(t)\nonumber\\
 a_{2,\vec k}(t)=b_{-,\vec k}(t)\beta_{-,\vec k}(t)-b_{+,\vec k}(t)\beta_{+,\vec k}(t).
\end{eqnarray}

We start from the instantaneous ground state at $t=0$ and thus we have $a_{1,\vec k}(0)=\beta_{+,\vec k}(0)$
and $a_{2,\vec k}(0)=\beta_{-,\vec k}(0)$. At the final time $T$, we obtain $b_{+,\vec k}(T)=
a_{1,\vec k}(T)\beta_{-,\vec k}(T)-a_{2,\vec k}(T)\beta_{+,\vec k}(T)$.
The excitation or defect density ($n_d(T_f)$) at time $T_f$ is given by the sum of probabilities for being at the excited adiabatic levels, $i.e.$,
\begin{equation}
n_d(T_f)=\sum_k |b_{+,\vec k}(T_f)|^2.
\label{eq1}
\end{equation}

\begin{figure}
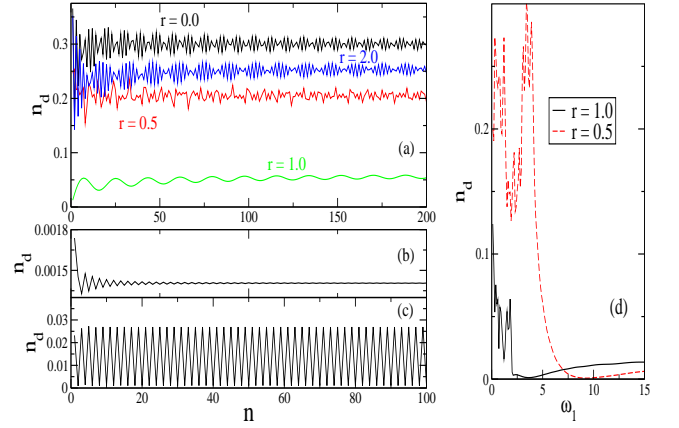

\vspace{.1in}
\centering
\includegraphics[width=.65\linewidth,height=2.2 in]{manycycle0.eps}\hskip .1 in
\includegraphics[width=.3\linewidth,height=2.2 in]{manycycle1b.eps}
\caption{(Color online) Plot of the defect density $n_d$ of the 1D XY model in
a transverse field as a function of number of drive cycle $n$ for
(a) drive frequency $\omega_1~=~1$ with $r~=$ 0, 0.5, 1.0 and 2.0. 
(b) drive frequency $\omega_1~=~4$ with $r~=$ 1.
(c) drive frequency $\omega_1~=~10$ with $r~=$ 0.5.
Dynamic freezing observed in (b) and (c) can also be seen from the $n_d$ versus $\omega_1$ 
plot in (d) for $n~=~20$.
Here we have chosen $J_x+J_y=1$, $J_x-J_y = \cos(r
\omega_1 t)$, and $h= A\cos(\omega_1 t)$ with $A=4$.}
\label{fig1}
\end{figure}
\begin{figure}[t]
\vspace{-.1in}
\centering
\includegraphics[width=.33\linewidth,height=2.2in]{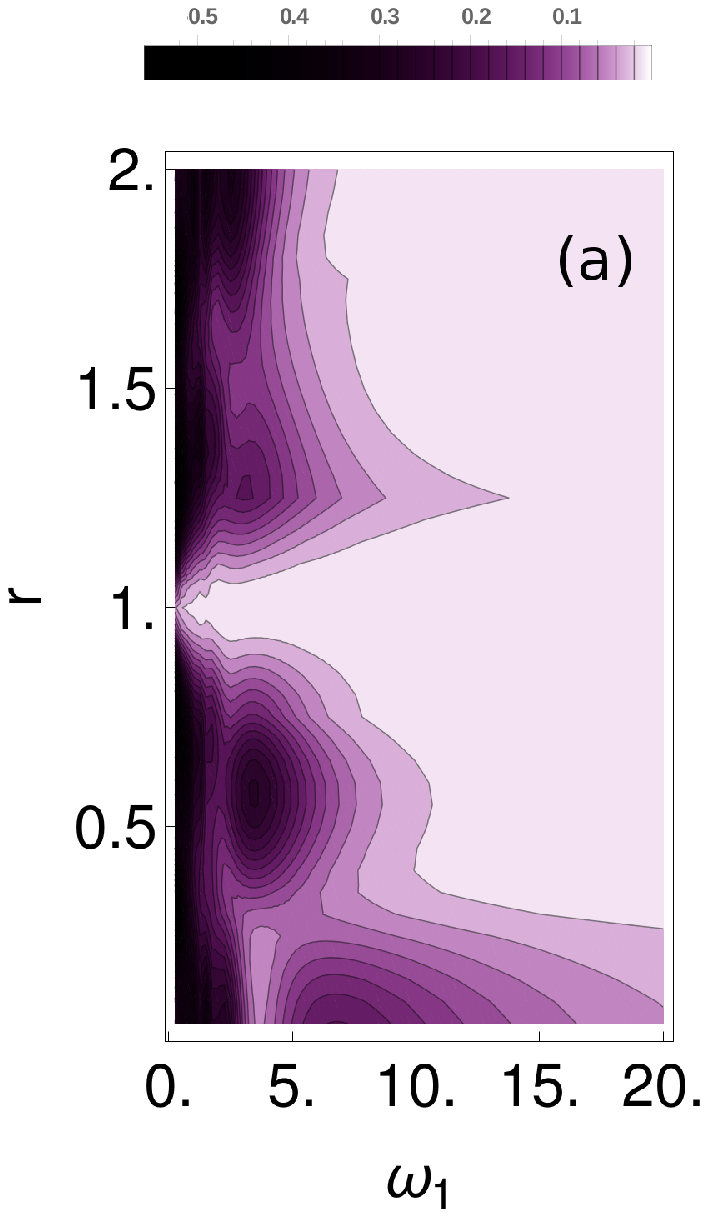}
\includegraphics[width=.33\linewidth,height=2.2in]{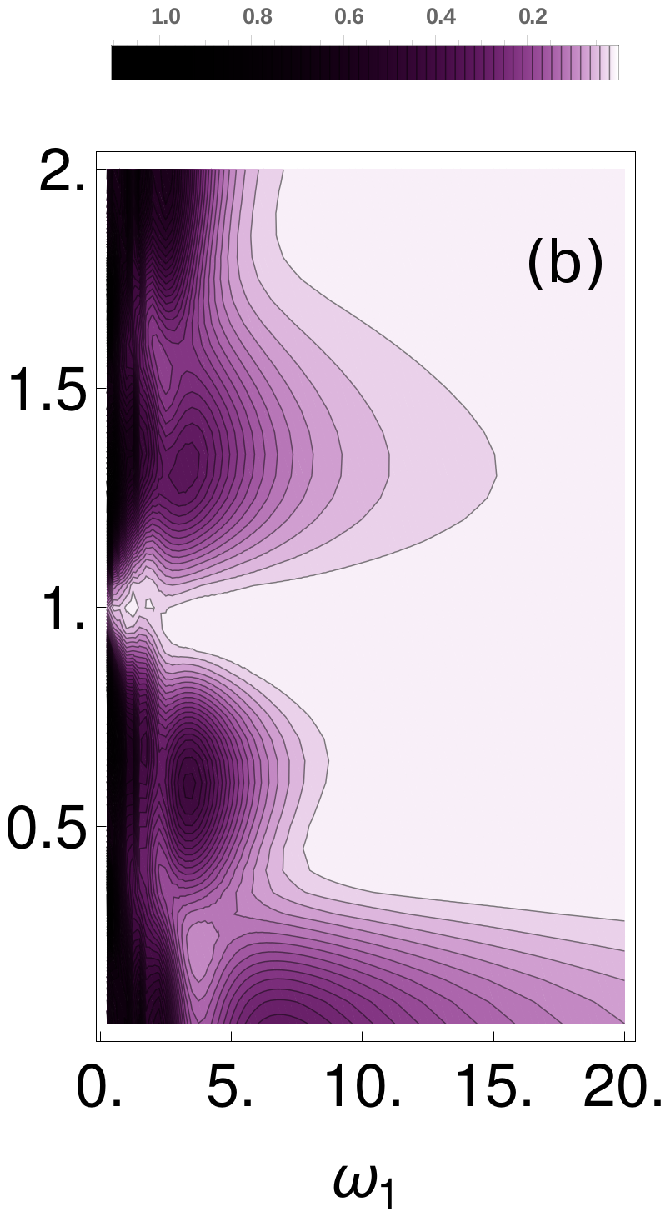}\hskip -.01 in
\includegraphics[width=.33\linewidth,height=2.2in]{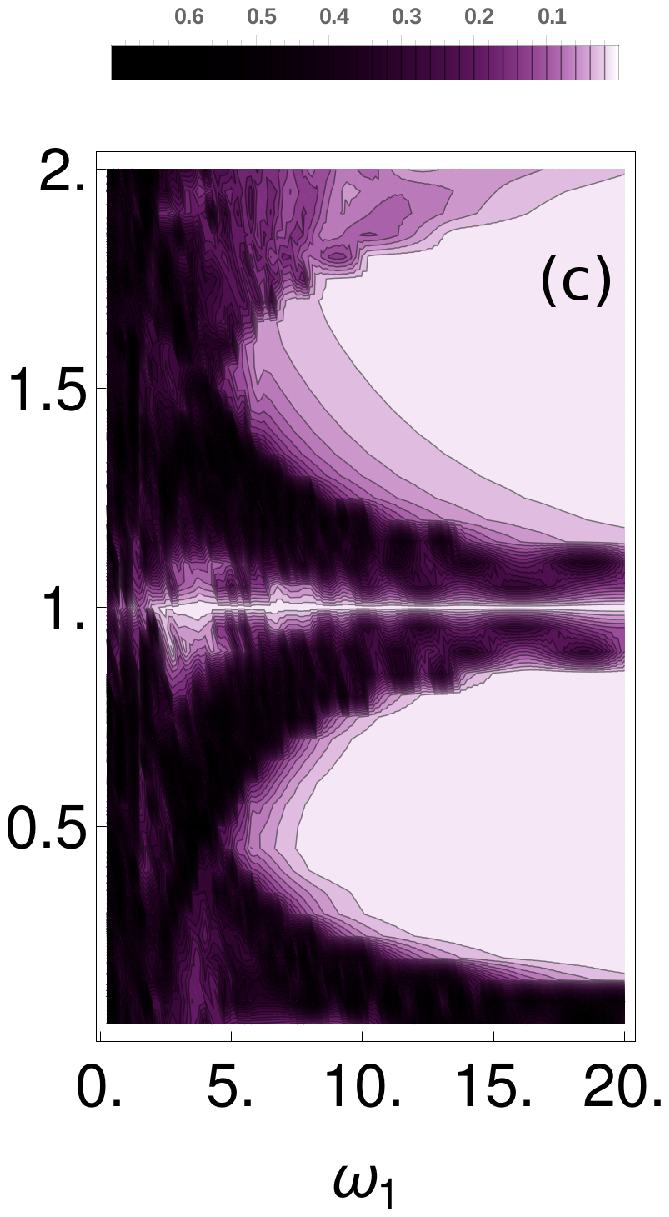}
\caption{
(Color online) Results from 1D TFXYM. (a) Defect density profile and (b) magnitude of the change in magnetization $(\Delta m_z)$ within a 
($\omega_1,r$) window after driving through $n=1$ cycles. (c) $\Delta m_z$ for $n=100$.
All parameters are same as in Fig.\ \ref{fig1}.}
\label{fig2}
\end{figure}

Our results are obtained mainly on a 1D TFXYM.
For that we first choose the parameters $\lambda_{1(2)}$, $b_{\vec
k}$, and $\Delta_{\vec k}$ to be those for a spin-$\frac{1}{2}$ XY chain in a transverse
field (Eq.\ \ref{mapxy}) with $\lambda_1 = -h = A \cos(\omega_1 t)$ and
$\lambda_2= \cos(r \omega_1 t)$. Furthermore we consider $J_{x(y)}= (1+(-)\lambda_2)/2$. While discussing $XY$ model throughout this paper, we consider a fixed amplitude $A=4$, the value same as that was used 
in Ref.\onlinecite{kar} to produce their main results.
With driving for long enough, we witness $n_d(T_f)$ to converge to some steady values (see Fig. \ref{fig1}(a)).
All our measurements are stroboscopic, in terms of the periodic bias signal having frequency $\omega_1$ 
($i.e.$, the time difference between the initial and final states are given by $nT=n\frac{2\pi}{\omega_1},~n$ being an integer).
Ref.\onlinecite{kar} shows that dynamic freezing or almost complete  overlap between initial and final wave-functions
occurs nicely with the two-rate protocol, 
mostly for $r=1$, when observed after one cycle of driving with frequency $\omega_1$. 
A scan through higher frequency regime reveals that such freezing zones grows with driving frequency, even sprouting
new freezing patches around $r\sim0.5$ and $2.0$ (within the regions shown in Fig. \ref{fig2}) for $\omega_1\sim 10$ and beyond.
We should mention here that freezing increases for very small and very large frequencies 
as $|b_{+,\vec k}(T_f)|\rightarrow0$ for all $\vec k$ values in those situations. This can be reasoned nicely
 using an adiabatic-impulse model to be discussed in the next sub-section.
Long stroboscopic driving smoothens out the temporal fluctuations in transient times and causes
freezing phenomena to go rarer.

  Fig.1(a) shows defect densities in a transverse field XY chain
for different r values as a function of driving cycle number $n$ for a frequency $\omega_1=1$.
Defect production tends to saturate as $n$ is increased gradually. The fluctuations above the steady value seems to have a period
which is multiple of $2\pi/\omega_1$, if $r$ is zero or an integer.
We see that the plot for $r=1$ shows little defect productions compared to that shown for other $r$ values as
the transition probability for a single passage through QCP for $r=1$ can be shown, in this case, to close to unity as compared to
its $r=0$ counterpart (also see Ref.\onlinecite{kar}).
The temporal evolution of $n_d$ is also more smooth for $r=1$.

We witness a low frequency freezing for $3\leq\omega_1\leq5$ for $r=1$ as well as a high or intermediate frequency freezing
for $r=0.5$ at $\omega_1\sim 10$. The corresponding plots can be seen in Fig.\ref{fig1}(b)-(d). Fig.\ref{fig1}(b),(c)
show the defect evolution with time for the corresponding cases (see that, 
Fig.\ref{fig1}(c) corresponds to $r=0.5$ and thus the overall periodicity is observed in units of 2 cycles) while Fig.\ref{fig1}(d)
demonstrates those freezing regimes for $n = 20$.
 Fig.\ref{fig2}(a) shows the defect density profile in the frequency space for $n = 1$.
An ultra-high frequency freezing is a common phenomena  for measurements done stroboscopically.
But the interesting thing to notice here is the dependence of the low $\omega_1$ cut-off, so to say,
on $\omega_2$ values, beyond which freezing continues as $\omega_1$ is increased further.
 A perfect freezing implies an unaltered state and thus any state variable will retain its values at freezing.
 The transverse magnetization $m_z(t)$ in the driven state can be defined as
\begin{eqnarray}
m_z (t) &=& 2 L^{-d} \sum_{\vec k} |a_{1\vec k}(t)|^2-1.
\end{eqnarray}
It thus gives a measure of occupation probabilities of the wave-function in the diabatic basis and its change between initial and final time
indicates a state change.
In Fig.\ref{fig2}(b)-(c), the magnitudes of the change in $m_z(t)$ (let's call it $\Delta m_z$) are shown for dynamics after single cycle as well as
after 100 cycles corresponding to $\omega_1$.
As freezing implies complete or almost complete overlap of the initial and final wave-functions, the change in $m_z(t)$ should
tend to zero under such scenario. Here we demonstrate that from the similarities of the regions $\Delta m_z\sim 0$
in Fig.\ref{fig2}(b) with the regimes of $n_d\sim 0$ in Fig.\ref{fig2}(a). As with long driving, the system tends towards the steady state,
the transient freezing behaviors get rarer as can be seen from Fig.\ref{fig2}(c). For example, The freezing points
of the $(\omega_1,r)$ plane, shown in Fig.\ref{fig1}(d) for $n=20$ existed within the huge
 freezing zone of Fig.\ref{fig2}(a),(b) at $n=1$, continues to exist at $n=100$ in Fig.\ref{fig2}(c). 
 However the freezing seen around $r\sim2$ for $\omega_1\sim10$ at $n=1$
 falls short at $n=100$, or even much before at $n=20$.
\begin{figure}
\vspace{-.1in}
\centering
\includegraphics[width=.48\linewidth,height=1.8in]{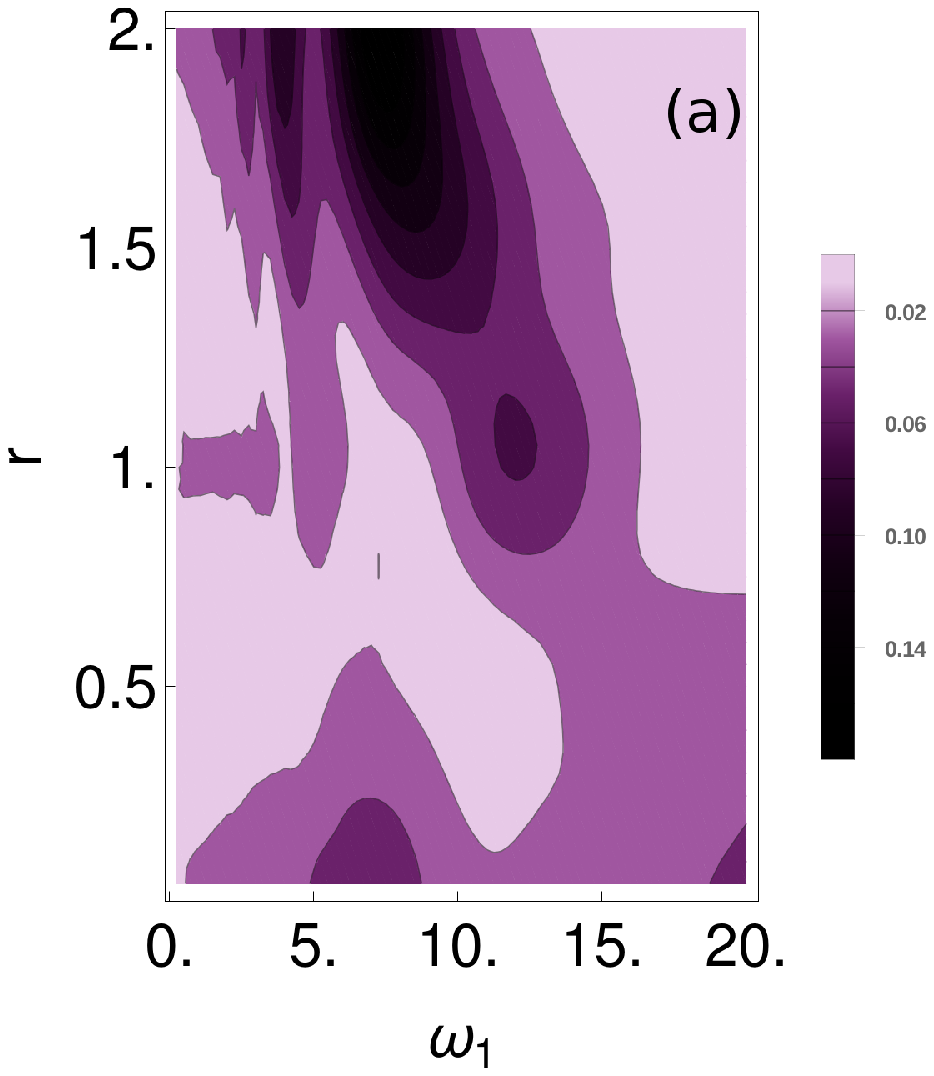}
\includegraphics[width=.48\linewidth,height=1.8in]{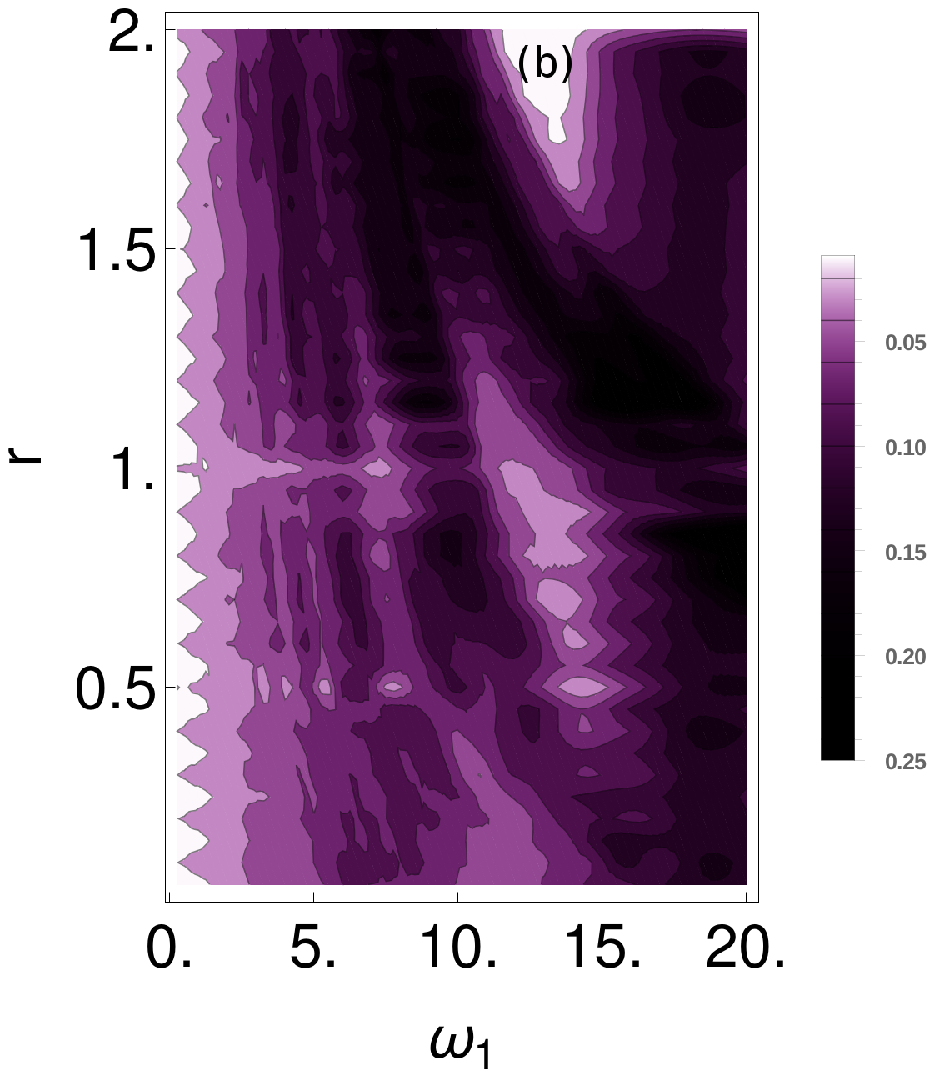}
\caption{
(Color online) Defect density $n_d$ in a 2D Kitaev model within a ($\omega_1,r$) window after driving 
through (a) $n=1$ and (b) $n=10$ cycles.
All parameter values are given in subsection IIA1.}
\label{fig3}
\end{figure}

The freezing region around $r=1$ is already known to appear for a driving through a single cycle\cite{kar} (Fig\ref{fig2}(b)
of this paper and Fig.5 of Ref.\onlinecite{kar} can be compared in this regard).
And here we show that freezing starts developing for other $r$ values as well when a larger range of $\omega_1$ values are 
considered. Moreover the freezing condition remains intact as $\omega_1$ is increased further.
However, after a long drive corresponding to $n=100$, the freezing scenario wears off as in a long time evolution
the evolved state spreads out more within the Hilbert space leaving less chances for the final state to be close to the initial state in the 
phase space. Fig.\ref{fig2}(c) shows regime 
about $r=0.5$ and $r=1.5$ where $\Delta m_z$ is small, but it is still larger compared to that of the freezing zones observed at $n = 1$.
We also notice that the range of variation in $\Delta m_z$ reduces as we compare the results for $n=1$ with $n=100$,
which indicates that after long time, the dynamics relaxes within a low-energy sector of states where magnetization 
does not vary much.

We also show in Fig.\ref{fig3} defect density profile in the $(\omega_1,r)$ space for 2D Kitaev model on a honeycomb lattice with $J_1=J_2=[1+cos(\omega_2t)]$
and $J_3=4.5[2+cos(\omega_1t)]$, as in Eq.\ref{kiteq}. It also demonstrate the same feature of reduction of dynamic
freezing as system is exposed to long driving. One distinctive feature that we can readily see here is that unlike
XY chain results, the 2D Kitaev model defect density profile shows pockets of dynamic freezing in 
the $(\omega_1,r)$ plane. This can happen due to the 
stuckelberg interference\cite{nori1} to be discussed in the next sub-section.

\subsubsection{Analytic approaches}

It is always good to have a theoretical handle of the results obtained using numerical integration 
of the Schrodinger equation in any dynamical problem. In what follows we will discuss the adiabatic-impulse
approximation (AIA) and rotating wave approximation (RWA) for our drive with the two rate protocol
to see how well they captures the dynamics at small and large frequency regimes respectively.
\\
\\
\textbf{Adiabatic-Impulse approximation}\\

The key aspect of the adiabatic-impulse approximation (AIA) is to
divide the dynamics of a system subjected to a drive into two
distinct regimes \cite{nori1}. The first is called the adiabatic regime
where the rate at which the system Hamiltonian changes with time is small
compared to the instantaneous energy gap; here the dynamics merely
gathers a phase of the system wave-function. The
second constitute the impulse regime where the rate of change of the
Hamiltonian parameter is comparable to or larger than the
instantaneous energy gap; in this regime, excitations are produced
since the system can no longer follow the instantaneous ground
state. In the context of the Hamiltonian given by Eq.\
\ref{fermhamden}, the latter region occurs when the system reaches a
QCP. Therefore AIA becomes accurate
for low-frequency drives where the small driving velocity forbids excitations to
occur in the adiabatic regimes leaving the impulse regimes alone responsible
for the defect productions.

The adiabatic-impulse formalism for a two rate protocol has been discussed
at length in Ref.\onlinecite{kar} where periodic driving up to single cycle is studied.
For the sake of continuity to the reader we here briefly outline the formalism before
discussing the extension of it for many cycles of driving.
To treat the dynamics of Eq.\ \ref{fermhamden} subjected to a
two-rate periodic protocol using the adiabatic-impulse
approximation, we first identify the adiabatic and the impulse regions
 and the critical points $t_{1(2){\vec k}}$ within a single cycle.
\begin{figure}
\vspace{.1in}
\centering
\includegraphics[width=\linewidth,height=1.8in]{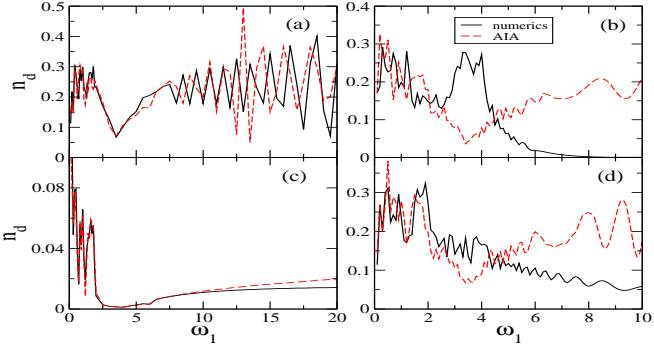}
\caption{(Color online) Comparison of exact numerical result (black solid line) for
variation of $n_d$ after $n=100$ cycles as a function of $\omega_1$ with that obtained
from the adiabatic-impulse method (red dashed line) for (a) $r=0$, (b) $r=0.5$, (c) $r=1$ and (d)
$r=2$. All parameters are same as in Fig.\ \ref{fig1}.}
\label{fig4}
\end{figure}

The gap, $2E_{+,\vec k}$ (see Eq.\ref{gap}) reaches a minimum at $t=t_{1,2 \vec k}$ at an 
avoided level crossing\cite{kar} for which $\partial E_{+,\vec k}/\partial t =0$. 
Around $t=t_{1,2 \vec k}$, the system enters the impulse regions and
excitations are produced. One important aspect of AIA
is to approximate the impulse region to be exactly at $t_{1,2, \vec k}$. 
This is obtained by linearizing
the Hamiltonian around $t_{1,2, \vec k}$. An unitary transformation can be engineered to give the
Hamiltonian at/ around impulse point  
a Landau-Zener (LZ) like form where linear time dependence appears in either diagonal or off-diagonal
part of the Hamiltonian. The reader is referred to Ref.\onlinecite{kar} for the detailed calculations. The transformations finally
give the Hamiltonian a LZ-like form whereby the transition probability at the impulse point $t_{a,\vec k}$ takes the form, 
\begin{align}
&~~~~~~~~~~~~~p_{\vec k}^a={\rm exp}[-2\pi\delta_{\vec k}^a]~~~~~{\rm with}~~~~~~~~~~\delta_{\vec k}^a=\nonumber\\
&\frac{((\lambda_1(\omega_1 t_{a,\vec k})-b_{\vec k})\dot \lambda_2(\omega_2 t_{a,\vec k})
-\lambda_2(\omega_2 t_{a,\vec k})\dot \lambda_1(\omega_1 t_{a,\vec k}))^2\Delta_{\vec k}^2}
{((\dot\lambda_1(\omega_1 t_{a,\vec k}))^2+\dot \lambda_2(\omega_2 t_{a,\vec k})^2\Delta_{\vec k}^2)^{3/2}}
\label{pk}
\end{align}

 With these and evaluating the the adiabatic evolution matrices between the critical points, we can reach the
 defect production probability $n_d(T)$ following Eq.\ref{eq1} with
\begin{align}
 &|b_{+,\vec k}|^2=p_{\vec k}^1(1-p_{\vec k}^2)+p_{\vec k}^2(1-p_{\vec k}^1)-\nonumber\\
 &2\sqrt{p_{\vec k}^1 p_{\vec k}^2(1-p_{\vec k}^1)(1-p_{\vec k}^2)}{\rm cos}(\theta_1+\theta_2+2\xi_2)=P_{\vec k}(T).
 \label{bk+}
\end{align}
Here $P_{\vec k}(T)$ is the excitation probability  in the ${\vec k}$-th mode.
The Stokes phase\cite{nori2,nori1,bhaskar}
$\theta_a=\frac{\pi}{4}+\delta_{\vec k}^a (ln \delta_{\vec k}^a-1)+{\rm arg}[{\Gamma} (1-i\delta_{\vec k}^a)]$ and
$\xi_1(\xi_2)=\int_{t_{1,\vec k}(t_{2,\vec k})}^{t_{2,\vec k}(t_{1,\vec k}+2\pi/\omega_1)}E_{+,\vec k}(t)dt$. 
For $n$ number of cycles, the transition probability is given by the expression\cite{nori1}
\begin{align}
&P_{\vec k}(nT)=P_{\vec k}(T)\frac{{\rm sin}^2n\phi}{{\rm sin}^2\phi}~~~~~~~~{\rm where}\nonumber\\
&~{\rm ~~cos}\phi=-\sqrt{(1-p_{\vec k}^1)(1-p_{\vec k}^2)}{\rm cos}(\theta_1+\theta_2+\xi_1+\xi_2)\nonumber\\
&~~~~~~~~~~~~~~~~~~~~~~~-\sqrt{p_{\vec k}^1 p_{\vec k}^2}{\rm cos}(\xi_1-\xi_2).
\end{align}
Within the adiabatic-impulse model, we can try to understand the freezing phenomena for driving up to, say,
single cycle only. Following AIA, the transition between the adiabatic states can occur only at the QCP or avoided
crossing points. With $\theta_1=\theta_2=\theta$ and $p_k^1=p_k^2=p_k$, say, for simplicity within Eq.\ref{bk+},
we obtain $|b_{+\vec k}|^2=4p_k(1-p_k)cos^2\Theta$ where $\Theta=2(\theta+\xi_2-\pi/2)$. For freezing this quantity 
needs to be zero or very small for all $\vec k$ values. Now $p_k=exp(-2\pi\delta_k)$, as Eq.\ref{pk} suggests. 
With very small or very large $\omega_1$ values, $\delta_k$ becomes very large or small 
causing $p_k$ to tend to 0 or 1 respectively.
In either case, this makes $|b_{+\vec k}|\rightarrow 0$ causing freezing. 
Moreover, $\delta_k$ also depends on the second frequency $\omega_2$.
So how fast $p_k\rightarrow1$ for high frequencies also depends on $\omega_2$. That's why we see a pattern for
lower cut-off of $\omega_1$ in the XY chain results in Fig.\ref{fig2}(b),(c) beyond which freezing continues
at higher frequencies. So far we have not considered the effect of angle $\Theta$ in this discussion.
But in principle, cos$\Theta$ can also be zero or very small enforcing the freezing criteria. This gives rise to the stuckelberg
interference\cite{nori1} and we may obtain pockets of dynamical freezing regime in the frequency space. 
2D Kitaev model results display similar features in its defect density profile in Fig.\ref{fig3}.

Comparison between exact numerical results and AI approximation can be seen in Fig.\ref{fig4}.
The defect density can nicely be reproduced using 
Adiabatic-Impulse approximation even after driving through many cycles. For small frequency the match is excellent while
it deviates as higher and higher frequencies are considered. It gives an estimate of a higher cut-off of the small
frequencies $\omega_1$ up to which AIA holds good.
Also we should point out here that the match is better for $r=0$ and 1 
as the Fig.\ref{fig4} demonstrates. This is because for $\omega_1\ne\omega_2\ne0$,
the time difference $n\frac{2\pi}{\omega_1}$ is not truly stroboscopic in nature and this fact need to be more suitably accommodated
in this AIA formalism.
\begin{figure}[t]
\vspace{.1in}
\centering
\includegraphics[width=\linewidth,height=2.4in]{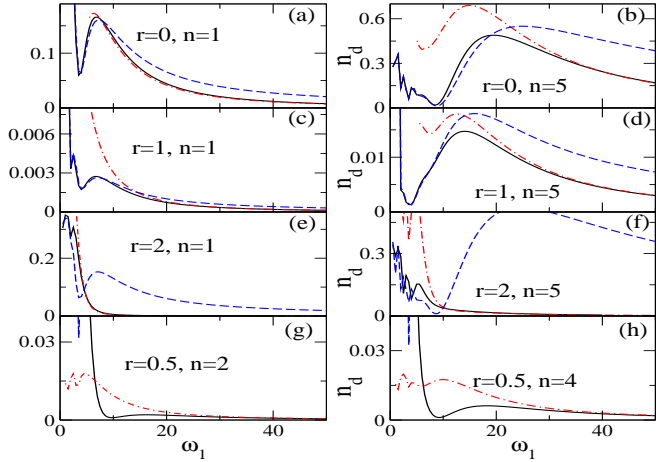}
\caption{(Color online) Comparison of exact numerical result (black solid line) for
variation of $n_d$ as a function of $\omega_1$ with that obtained
from the rotating wave approximation (red dash-dotted line) and adiabatic-impulse method (blue dashed line)
for (a) $r=0,~n=1$, (b) $r=0,~n=5$, (c) $r=1,~n=1$, (d) $r=1,~n=5$,
(e) $r=2,~n=1$, (f) $r=2,~n=5$, (g) $r=0.5,~n=2$, (h) $r=0.5,~n=4$. All parameters are same as in Fig.\ref{fig1}.}
\label{fig5}
\end{figure}
\\
\\
\textbf{Rotating wave approximation}\\

A rotating wave approximation (RWA) is an widely used approximation scheme in Quantum Optics and is often used to 
derive Rabi oscillations in a TLS.
Within the formalism of rotating wave approximation (RWA), a transformation to the interaction picture or 
a rotating frame is made\cite{nori2,nori1} via an unitary transformation, which in our case is
given by
\begin{align}
|\psi_{\vec k}(t)>&=U_{\vec k}(t)|\psi'_{\vec k}(t)>~~~{\rm with}\nonumber\\ U_{\vec k}(t)&={\rm exp}[-\frac{i}{2}(\eta-b_{\vec k}t)\tau_3]
~~~{\rm and}~~~\eta=\int \lambda_1dt~.
\end{align}
The objective, firstly, is to acquire a transformed $2\times 2$ Hamiltonian $H'_{\vec k}(t)$ 
that possess zero diagonal entries.  In the remaining time-periodic off-diagonal 
part, a power series expansion is utilized\cite{nori1} following the relation 
\begin{equation}e^{iz{\rm sin}\tau}=\sum_{n=-\infty}^\infty J_n(z)e^{in\tau}\nonumber
 \end{equation}
 where  $J_n(z)$ denotes the Bessel's function of first kind with order $n$ and argument $z$. The approximation comes when only the slowest 
 moving component of the periodic expansion is kept in the Hamltonian off-diagonal entries and thus it works well in high frequency limit. 
 Thereafter a further rotation from basis $|\psi'_{\vec k}>$ to $|\psi''_{\vec k}>$=
exp$[\frac{ib_{\vec k}t}{2}\tau_3]|\psi'_{\vec k}>$ is performed that results in a time independent Hamiltonian 
\begin{displaymath}
H''_{\vec k}=\left(\begin{array}{ccc}
-b_{\vec k}/2 & d({\vec k})/2 \\
 d({\vec k})/2 & b_{\vec k}/2
\end{array}\right).
\end{displaymath}
Here $d({\vec k})==\Delta_{\vec k}(J_m(A/\omega_1)+J_{-m}(A/\omega_1))/2$ for $r=m$, an integer. We should mention here that
the final states for an integral $r^{-1}$ can also be computed in a similar fashion at truly stroboscopic $n\tau$ times apart from the initial point,
where $n$ is an integer and $\tau$ is the lowest common multiple of 
$\frac{2\pi}{\omega_1}$ and $\frac{2\pi}{\omega_2}$ (See, for example, results in Fig. \ref{fig5}(g),(h)). This is because the diagonal and 
off-diagonal entries of $H_{\vec k}$ can be swapped within a transformation $H\rightarrow e^{-\frac{i\pi\tau_2}{4}}He^{\frac{i\pi\tau_2}{4}}$ and RWA 
can be performed over the transformed Hamiltonian.

 At this stage, it is easy to solve the Schrodinger equation and obtain the transformed states. Back transformations from there can
 retrieve the original states and we can express the time evolution of states as
\begin{equation}
|\psi_{\vec k}(t)>={\rm exp}[-i\frac{d({\vec k})t}{2}\tau_1]{\rm exp}[\frac{i}{2}(b_{\vec k}t-\eta)\tau_3]|\psi_{\vec k}(0)>.
\label{RWpsi}
\end{equation}

The formalism thus comes in steps, each of which involves a rotation of the wave-function basis. The approximation comes when the time 
dependent periodic function, which can be written 
as a sum over Bessel's functions times exponentials, is replaced with only the smallest (if not resonant) term present there.
Naturally this is an approximation good at high frequencies as the other components become increasingly fluctuating with higher frequencies.
We should mention here that RWA works much better if we look at time averaged variables after long times ($e.g.$, dynamical order parameter $Q$
as calculated in Ref.\onlinecite{adas1})
as the highly fluctuating components can average out to zero. However for an observable, measured instantly at a time far from the initial time, the mutual cancellation
of the number of highly fluctuating components give a good estimate as well.

Our RWA results, in comparison with exact numerics and AIA, are shown in Fig.\ref{fig5}.
The defect densities obtained using RWA can be seen to match with the exact results at high frequencies.
We also notice that the comparison with RWA becomes better for longer times of driving if larger values of $\omega_1$ are considered.
To explain that we need to see that RWA is exact for $\omega_1\rightarrow\infty$ and is a better approximation for larger $\omega_1$.
But at the same time, approximation is used in obtaining the expression $d({\vec k})$ in the transformed Hamiltonian $H''_{\vec k}$. 
So the error in an instantaneous state measurement increases with longer $t$ because
of the factor $d({\vec k})t$ appearing in the expression of the evolved wave-function (Eq. \ref{RWpsi}).

\subsection{Entanglement generation}

The Schrodinger time evolution of the ground state in a closed quantum many body system does not carry any entropy with it.
Nevertheless, the states can get entangled and we can attach an entanglement entropy to the state as an estimate for such
entanglement.
The entanglement entropy $S_n(l)$ of a sub-system of size $l$ within our free fermionic driven system
can be obtained from the two-point fermionic correlators $C_{ij}$ and
$F_{ij}$:
\begin{eqnarray}
 C_{{i} {j}} &=& \frac{2}{ L^d} \sum_{\vec k\epsilon BZ/2} |a_{1, \vec k}(t)|^2
\cos[\vec k \cdot (\vec i - \vec j)] \nonumber\\
F_{ij} &=&  \frac{2}{ L^d} \sum_{\vec k\epsilon BZ/2} a_{1, \vec k}^\star(t)a_{2, \vec k}(t)
\sin[\vec k \cdot (\vec i - \vec j)].
\end{eqnarray}
 A block diagonal $2l^d\times2l^d$ correlation matrix $\mathcal{C}$ is constructed with $l\times l$ diagonal blocks
$1-C$ and $C$ and off-diagonal blocks $F$ and $F^\star$ respectively and then diagonalized
to obtain the eigenvalues $p_i$'s\cite{sen1,marino}. From there the entropy is obtained as $S_n(l)=-\sum_1^{2l}p_i{\rm log}(p_i)$. 

\begin{figure}
\vspace{.1in}
\centering
\includegraphics[width=.9\linewidth,height=1.5in]{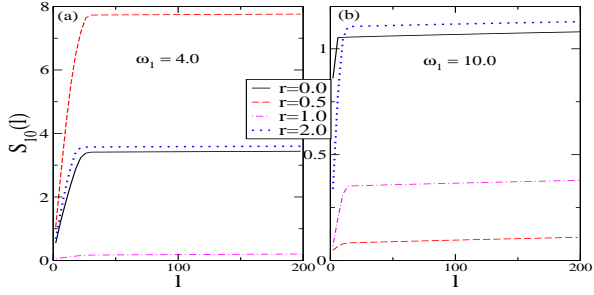}
\caption{(Color online) Entanglement entropy $S_n(l)$ after $n=10$ cycles as a function of sub-system size $l$
for different values of $r$ at (a) $\omega_1=4.0$ and (b) $\omega_1=10.0$. Parameters used are same as in Fig.\ref{fig1}.}
\label{fig6}
\end{figure}

Let us first look into any possible connection between entanglement and freezing. A perfect freezing keeps the state intact and hence
the amount of entanglement remains the same. The initial state is the the many-body ground state, which for short-ranged 
Hamiltonian, within the gapped phases, generally remains unentangled.
Thus the region of maximum freezing should appear with minimum values for $S_n(l)$ after the dynamics.
That is exactly what is seen in Fig.\ref{fig6}(a)-(b), for the freezing points depicted as in Fig.\ref{fig1}(d). 
The near-zero entanglement estimates appear at points in $(\omega_1,r)$ plane, as mentioned early in section II-A-1.
\begin{figure}
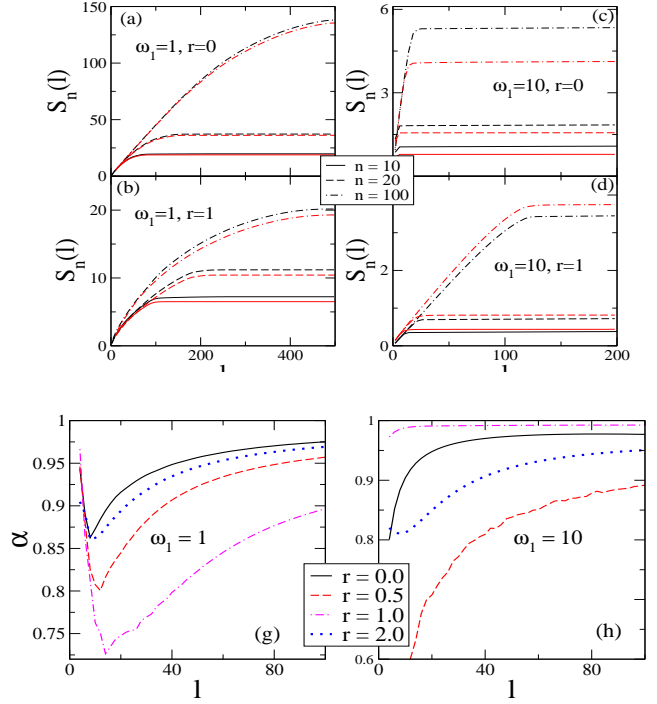

\vspace{.1in}
\centering
\includegraphics[width=.9\linewidth,height=2.in]{Sn4b.eps}\\
\vskip .15 in
\hskip -.1 in
\includegraphics[width=.98\linewidth,height=1.5in]{Sn5d.eps}
\caption{(Color online) $S_n(l)$ versus $l$ plots for $n$=10(solid line), 20(dashed) and 100(dash-dotted) at (a) $\omega_1=1.0,~r=0$, (b) $\omega_1=1.0,~r=1$, 
(c) $\omega_1=10.0,~r=0$, (d) $\omega_1=10.0,~r=1$. Plots obtained within AIA are shown (red/gray lines) demonstrating
better matches at smaller frequency $\omega_1=1$.
Exponent $\alpha$ versus $l$ for various $r$ values with $n=1000$ at (e) $\omega_1=1$ and (f) $\omega_1=10$ ($n=10000$ is used
for $r=0$ here). All parameters are same as in Fig.\ref{fig1}. }
\label{fig7}
\end{figure}

Next we turn to a more general discussion on $S_n(l)$ in our system.
According to Hastings' theorem\cite{hastings}, a system with short range interactions, like a TFIM or TFXYM, should follow an area scaling for its 
entanglement entropy, $i.e.$, $S\sim l^{d-1}$ at dimensionality $d=1$.  
For a periodic drive, we indeed find such behavior at $l\sim L/2$, starting from a cut-off $l_c$ that increases with $n$, the number of 
drive cycles. For very large $n$, we get $l_c\sim O(L/2)$ and thus most of the time the entropy follows a non-area law of scaling. In fact it 
follows a volume law: $S\sim l^d$, for $n\rightarrow\infty$. 

The results for $S_n(l)$ for a 1D TFXYM are shown in Fig.\ref{fig7}.
The entanglement entropies of subsystems A and B of a composite system in a pure state
are equal. No entanglement can be there for a subsystem of null size. Hence we find  $S(l=0)=S(l=L)=0$.
The area law sprouts from the fact that bipartite entanglement in the ground state of a local Hamiltonian of a gapped system is proportional  
only to the boundary area between the two subsystems and hence it obeys: $S\sim l^{d-1}$. Now for very small $l$, the whole volume of
the small subsystem contributes to entanglement, irrespective of whether the Hamiltonian is short-ranged or not. So we expect a volume law 
scaling for $S$ at $l<<L$. But as $l$ increases, so does the volume-to-area ratio for the subsystem and the entanglement tends to
follow the area scaling more and more. As the subsystem size exceeds the correlation length $l_0$ (which is of the order of 
the cut-off $l_c$) of the correlators 
$C_{ij}$ and $F_{ij}$, area scaling appears to dominate in the $S_n(l)$ behavior (see Fig.\ref{fig7}(a)-(d)). 
Thus we see a gradual cross-over from the volume scaling to area scaling behavior for $S(l)$ as $l$ is increased from 0 up to $L/2$ with
continuous regime of non-area-non-volume scaling behaviors in between. A large correlation in the time evolved state, however, pushes $l_c$ 
more towards $L/2$. 
\begin{figure}
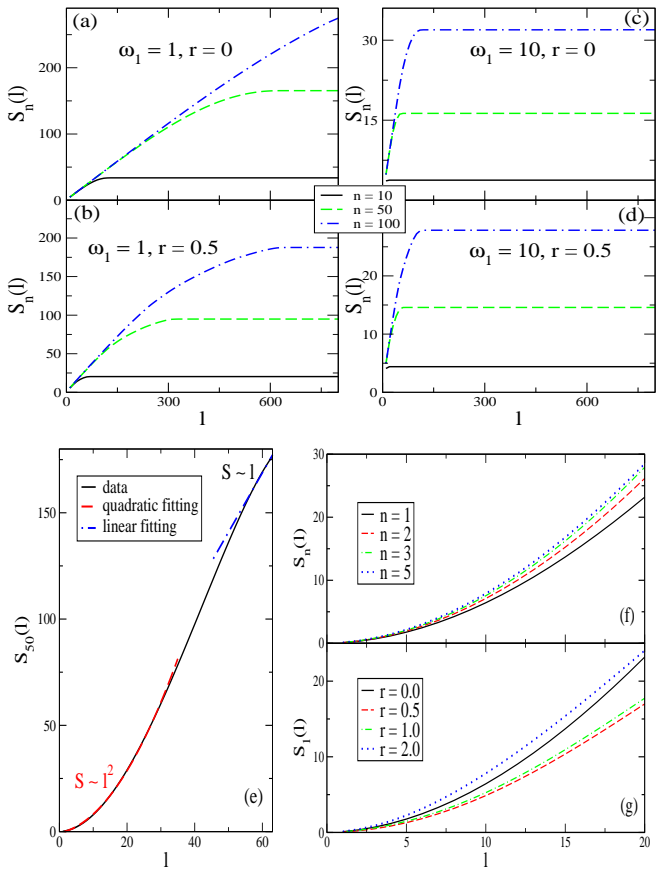

\vspace{.1in}
\centering
\includegraphics[width=\linewidth,height=2.2 in]{kitaev1D.eps}\\\vskip .1 in
\includegraphics[width=.4\linewidth,height=2.2 in]{n50L100kit2D.eps}\hskip .1 in
\includegraphics[width=.55\linewidth,height=2.2 in]{kitaev2D-n.eps}
\caption{(Color online) $S_n(l)$ versus $l$ plots for $n$=10(solid line), 50(dashed) and 100(dash-dotted) at (a) $\omega_1=1.0,~r=0$,
(b) $\omega_1=1.0,~r=0.5$, (c) $\omega_1=10.0,~r=0$, (d) $\omega_1=10.0,~r=0.5$ for a Kitaev chain with parameters mentioned in
section II-B. (e) Scaling of $S_{50}(l)$ with $l$ for a 2D Kitaev model honeycomb lattice. 
Variation of (f) $S_n(l)$ for different $n$ values at $r=0$ and (g) $S_1(l)$ for different $r$ values. A $100\times 100$ rhomboidal mess
is considered for the calculations in 2D Kitaev model. The parameter values are mentioned in section II-B.}
\label{fig8}
\end{figure}

In Fig.\ref{fig7}(a)-(d), we further see that for a fixed $(\omega_1,\omega_2)$ pair, the magnitude of $S(l)$ as well as the cut-off $l_c$ to increase with $n$. The ground 
state of the gapped many-body system at $t=0$ has very little entanglement and can thus be approximated as a product state.
Starting from there, as the state is dynamically evolved for a long time, its spread within the Hilbert space keeps increasing 
and it becomes more entangled. A large $l_c$ implies a larger subsystem size for
which boundary rather than the whole volume becomes relevant for entanglement. For large $n$, correlation in the system increases and we need 
to go to higher cut-off values where the smaller area-to-volume ratio can compensate for the larger correlations reaching out to area scaling
more.

For very small as well as very large frequencies, a stroboscopic dynamics hardly see the state to evolve from its starting ground state
and thus we obtain both freezing and unentangled final states.
A fast driving reduces the correlation length which is manifested here from lower $l_c$ values in Fig.\ref{fig7}(c),(d) 
compared to that in Fig.\ref{fig7}(a),(b) (as also in Fig.\ref{fig8}(c),(d) compared to Fig.\ref{fig8}(a),(b)
from a 1D Kitaev model results). Also notice that the small frequency $(\omega_1=1)$ entanglement entropy results of
Fig.\ref{fig7}(a),(b) can be well represented by Adiabatic-Impulse approximation (shown there as well), compared 
to similar results for large frequencies (such as $\omega_1=10$, as shown in Fig.\ref{fig7}(c),(d)), as AIA supposedly works nicely for slow
quenches.

If we define, for the entanglement entropy, an exponent of scaling\cite{sen1} as $S\sim l^\alpha$, we see that $\alpha$ tends to unity, in our TFXYM chain, if $l$ is gradually increased staring from zero 
or decreased starting from $L$. However this convergence towards unity is often non-monotonic. These results are demonstrated in Fig.\ref{fig7}(e)-(f). 
For small $n$, entropy follows the area scaling above some cut-off values of $l$. But this cut-off increases with the values of n. 
In the plateau region above the cut-off, we witness the area scaling law to hold. But the region before cut-off becomes 
larger with $n$ where we see the non-area, non-volume scaling laws to occur. For small $\omega_1$ (Fig.\ref{fig7}(a),(b)), the plateau region soon vanishes 
as $n$ becomes large whereas a large $\omega_1$ retains such behavior even for the large $n$ values shown here in Fig.\ref{fig7}(c)-(d). 
Fig.\ref{fig7}(e)-(f) shows the evolution of scaling exponent $\alpha$ with $l$ for a large $n=1000$ cycle of driving and demonstrate the convergence 
of $\alpha$ towards unity. For a low frequency such as $\omega_1=1$, the 
scaling exponent goes to 1 rather nontrivially as can be seen at the smallest end of the sub-system size $l$ (Fig.\ref{fig7}(e)). For large 
$\omega_1$, the behavior is mostly monotonic 
The convergence towards unity is rather slow for $r=0$ (see Fig.\ref{fig7}(f)).
So a larger $n=10000$ is used to see behavior of $\alpha$ in Fig.\ref{fig7}(f). We see that for $r=1$, $\alpha$ to converge to unity 
 faster with $l$ than other values of $r$ shown in Fig.\ref{fig7} if larger frequencies (such as $\omega_1=10$) are considered. For small frequencies 
 instead  ($\omega_1=1$), such convergence is slowest for $r=1$.
\begin{figure}
\vspace{.1in}
\centering
\includegraphics[width=\linewidth,height=2.5 in]{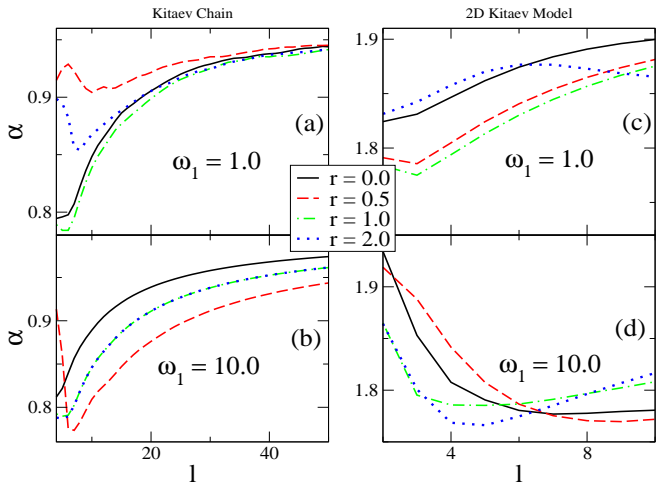}
\caption{(Color online) 
Exponent $\alpha$ versus $l$ at various $r$ values for (a),(b) Kitaev chain ($L=2000$) with $n=1000$ 
and (c),(d) 2D Kitaev model ($L=200$) with $n=100$. Results are shown for (a),(c) $\omega_1=1$ and (b),(d) $\omega_1=10$ .
Parameters used are those mentioned in section II-B.}
\label{fig9}
\end{figure}

Thereafter we also study the entanglement entropy behaviors in Kitaev chain and 2D Kitaev model on a honeycomb lattice.
We choose $J_1=0,~J_2=1+cos(\omega_2t),~J_3=[1+cos(\omega_1t)]/2$
for Kitaev chain and $J_1=J_2=[1+cos(\omega_2t)],~J_3=4.5[2+cos(\omega_1t)]$ for 2D Kitaev model so that
both represent gapped systems at initial and stroboscopically advanced final times.
The results from Kitaev chain, as shown in Fig.\ref{fig8}(a)-(d), are in essence similar to 1D TFXYM results of
Fig.\ref{fig7}(a)-(d). In the 2D Kitaev model case, however, the entanglement entropy results bear the signature of 
dimensionality 2. Here we consider $l\times l$ rhombus shaped sub-systems within a system of size $L\times L$. For each such sub-system, 
the environment or the other partition within the system has a non-rhombus shape
with area $L^2-l^2$ and thus, unlike in 1D, we generally have $S_n(l)\ne S_n(L-l)$. Starting from zero, $S_n(l)$ grows with $l$, reaches a maximum at an $l>L/2$
and then gradually decrease down to zero at $l=L$. 
For small $l$, volume scaling behavior is manifested and entropy scales as $S_n(l)\sim l^2$. As $l$ value is increased
the behavior gradually changes to a area law behavior with $S_n(l)\sim l$. 
Fig.\ref{fig8}(e) illustrates such behaviors.
We may point out here that had we considered a parallelogram shaped subsystem of 
size $l\times L$ ($i.e.,$ cylindrical subsystem with periodic boundary conditions) 
instead, we would get, like in 1D cases, $S_n(l)=S_n(L-l)$ with $S_n(l)$ 
becoming symmetric about $L/2$. We find that in such situation entropy scales
 as $S_n(l)\sim l^{1/2}$ for small $l$ values.

Similar to 1D model results, in 2D Kitaev model also
$S_n(l)$ is found to increase with $n$ (see Fig.\ref{fig8}(f)) and by tuning
$r$, we can tune the entropy as well. Fig.\ref{fig8}(g) shows variation in $S_n(l)$ for $n=1$ for different $r$ values.

Finally if we look at the plots of exponent $\alpha$ versus $l$ in Kitaev chain and 2D Kitaev model, we see that for 
very small $l$ values the behavior, in general, is non-monotonic. We can also compare these plots with 1D TFXYM
plots of Fig.\ref{fig7}(e),(f). When we say that for small $l$, we see an volume scaling law to hold that
gradually changes to a area scaling laws as $l$ is increased in these short ranged integrable models, we expect
the exponent $\alpha$ to start from $d$ at very small $l$ and to tend to $d-1$ when $l$ increases more and more.
However, we find that the $\alpha$ vs. $l$ plots for different $r$ at very small $l$ values ($i.e.,$ for $l<<L$) displays, rather, some
non-monotonic behavior.
This is certainly an interesting 
observation and deserves further investigation.

\section{Discussion}
\label{diss}

In this work, we have studied long-time periodic dynamics of a class of
integrable models using
a two-rate protocol. Our results on a 1D TFXYM reveals that, starting the dynamics from ground state of the time dependent Hamiltonian,
the periodic driving can lead to dynamical freezing when
the second drive frequency $\omega_2=r \omega_1$ is used as a tuning parameter.
Such freezing starts developing from pretty small values of $\omega_1\sim O(1)$
 when $r\sim 1$ and it starts spreading along $r$, particularly if the system is driven for a few cycles,
 as frequency is increased further. New freezing zones starts sprouting out at intermediate frequencies
 $\omega_1\sim O(10)$ for $r\sim0.5$ or $2.0$ and continues to grow similarly with higher $\omega_1$ values. 
 However, stroboscopic measurements after long driving sees the final state to get more and more steady with
 freezing scenario getting rarer compared to its transient profile. This is an expected trend
 and is independent of the models used, even though the distribution of the defect density in the frequency space
 varies, as confirmed from 2D Kitaev model results.
  
 In order to have a analytical framework of the said dynamics, we extend the formalism of adiabatic impulse approximation 
 for the two-rate protocol, as originally described in Ref.\onlinecite{kar},
to the case of driving up to multiple cycles and for higher frequencies. It shows near-exact comparison with exact results
at small frequencies (much higher than shown in Ref.\onlinecite{kar}, and for far beyond merely a single cycle).
We also develop a suitable extension of the rotating wave approximation scheme so that dynamics for fast quenching,
in this case, can also be accounted for analytically. Our extension works for integral $\omega_2/\omega_1$ or 
$\omega_1/\omega_2$ values.

 We also have a look at the entanglement entropy results in our dynamical system. We witness the unentangled nature of
 the ground states of our free fermionic Hamiltonians.
 The dynamic freezing regimes are found also to correspond to the regions of minimum entanglement in the system.
 Within the gapped phases, that we study, we find area scaling laws for entanglement entropy $S_n(l)$ to hold when
 driven not for long while genuine non-area scaling to occur under long driving. The scaling behavior tend to a volume scaling law for
 very large $n$. Within our two-rate protocol, a tuning with $r$ resulting in huge alteration of the entropy content
 can always be commercially used in our favor. Our results both on TFXYM and Kitaev chain and 2D Kitaev model on 
 honeycomb lattices  demonstrate important behavior of the periodically driven (with a two-rated protocol)
 integrable models away from a QCP.

Our work can be well extended to studying the long-lived non-thermal steady states\cite{pol1,steady}
when driving periodically with a two rate protocol. 
The dynamical freezing, obtained in this work, can motivate further researches  in the field of quantum computing,
such as designing superconducting qubits{\cite{Johansson} or circuit QED devices\cite{shev2}. 
Moreover, finding the Floquet spectrum and the possible change in its topology, already observed for $r=0$ case using a square 
wave protocol\cite{sen1}, can also give useful information on different dynamical phases and the transitions between them.
Furthermore, we may also look into the entanglement spectrum and Schmidt gap 
in the dynamically evolved states and probe the possibilities of any topological transition\cite{saptarshi} there.
In short, the tuning knob $r$  of the two rate periodic protocol opens up a world of opportunities enabling
exploration of multifarious dynamical phenomena with huge possibilities for numerous practical implementations.

{\it Acknowledgement}: The author benefited from discussions with K. Sengupta, H. K. Krishnamurthy, S. Mandal, 
A. Das, A. Sen and S. N. Shevchenko. 
This work is financially supported by CSIR, India, under Scientists' Pool Scheme No. 13(8764-A)/2015-Pool.

\end{document}